\documentclass[12pt]{article}
\usepackage{amsmath,amsfonts,amssymb}
\usepackage{graphicx}
\usepackage{setspace}
\usepackage{tocloft}
\usepackage{authblk}

\providecommand{\keywords}[1]{\textbf{\textit{Keywords---}} #1}

\begin{document} 
\title{Design and modelling of spectrographs with holographic gratings on freeform surfaces}

\author[a,b,*]{Eduard R. Muslimov}
\author[a]{Marc Ferrari}
\author[a]{Emmanuel Hugot}
\author[a]{Jean-Claude Bouret}
\author[c]{Coralie Neiner}
\author[a]{Simona Lombardo}
\author[a]{Gerard Lemaitre}
\author[a]{Robert Grange}
\author[b]{Ilya A. Guskov}
\affil[a]{Aix Marseille Univ, CNRS, CNES, LAM, Marseille, France}
\affil[b]{Kazan National Research Technical University named after A.N. Tupolev –KAI, 10 K. Marx, Kazan,Russia, 420111}
\affil[c]{LESIA, Observatoire de Paris, Université PSL, CNRS, Sorbonne Université, Univ. Paris Diderot, Sorbonne Paris Cité, 5 place Jules Janssen, 92195 Meudon, France}

\renewcommand{\cftdotsep}{\cftnodots}
\cftpagenumbersoff{figure}
\cftpagenumbersoff{table} 

\maketitle

\begin{abstract}
In the present paper we demonstrate the approach {of using} a holographic grating on a freeform surface for advanced spectrographs design. We discuss  the surface and groove pattern description used for ray-tracing. Moreover, we present a general procedure of diffraction efficiency calculation, which accounts for the change of hologram recording and operation conditions across the surface. The primary application of this approach is the optical design of the POLLUX spectropolarimeter for the LUVOR mission project where a freeform holographic grating operates simultaneously as a cross-disperser and a camera with high resolution and high dispersion.  The medium ultraviolet channel design of POLLUX is considered in detail as an example. Its resolving power reaches [126,000-133,000] in the region of 118.5-195 nm. Also, we show a possibility to use a similar element working in transmission to build an unobscured double-Schmidt spectrograph. The spectral resolving power reaches 4000 in the region 350-550 nm and remains stable along the slit.     
\end{abstract}



\begin{spacing}{2}   

\section{Introduction}
\label{sect:intro} 

The use of freeform optics in the field of optical technology is expanding the borders of achievable optical system performance.  With use of such surfaces it becomes possible to create wide-field fast optical systems with fewer number of optical elements and also create new systems with geometry, which impossible with ordinary aspheres.  
Application of a freeform optical surface for the  design of diffractive optical elements represents an attractive prospect. {Technology of complex diffractive optical elements, especially holographic ones, has been developing for a long time \cite{OShea2004, Palmer2005}, and the freeforms represent an emerging technology, which made a huge progress in the recent years \cite{Freeman2004, Burge2009, Fess2013}. Their combination could allow greater} freedom for aberrations correction and achieve a qualitatively new levels in spectroscopy, imaging, and beam shaping.
Such a possibility was recently considered and demonstrated by a number of authors. For example, an optical design with a blazed grating on a non-symmetrical tilted elliptical surface was developed and fabricated for the IGIS spectrograph~\cite{Bourgenot2016}.  In that case the grating had equally spaced straight grooves (in projection to the tangent plane).  Another freeform grating was used in the Offner-type design for the ELOIS spectrometer~\cite{Clercq2015}. It had a complex freeform shape without rotational symmetry and the groove pattern varying in order to maintain their positioning with respect to the local surface normal.  In general, application of freeform surfaces for the design of spectrographs was considered in different studies~\cite{Liu2017,Wei2016}. One of such publications~\cite{Reimers2017} demonstrates an Offner-Chrisp system with a grating on a general freeform surface described by Zernike polynomials. Also, there are ongoing research activities, which will allow {a combination of} freeform surface, complex groove pattern and blazed groove profile in a single optical element~\cite{Marchi2017}.  Finally, a number of publications have shown the {potential for application} of freeform diffractive elements in adjacent fields like display technologies, or image reconstruction ~\cite{LiuPang2017,LiuLiu2017}.
In the present paper we consider the most general case, when a grating with curved and unequally spaced grooves is imposed over a freeform surface. It is supposed that the grating is manufactured holographically, i.e. it represents a recording of interference pattern from two coherent sources. We developed a set of tools for ray-tracing through such a holographic freeform grating in Zemax and further computation of the diffraction efficiency. The primary application of these tools was the design of the POLLUX spectropolarimeter ~\cite{Bouret2018} for the LUVOIR ~\cite{Bolcar2017} mission concept. We show that the use of the freeform grating allows the required image quality {to be obtained} with the minimal number of optical surfaces and reach relatively high diffraction efficiency. However, the possible applications are not limited to this particular case. It is demonstrated that a transmission freeform grating can help achieve a high performance in a double-Schmidt type spectrograph. The {efficiency optimization} of a volume-phase hologram on a complex surface represents a separate task. 
Thus the paper is organized as follows: section 2 presents {how} to describe the freeform surface shape, the grating pattern and to model the diffraction efficiency accounting for the parameters variation across the surface; section 3 demonstrates the usage of such an optical element for the POLLUX instrument design with estimation of the obtained performance, including the image quality and the diffraction efficiency; section 4 presents a demonstrative optical design with a transmission freeform grating showing a similar performance assessment. Finally, section 5 contains the general conclusions on the study.

\section{Design approach}
\label{sect:design} 
Describing a holographic grating on a freeform surface requires {definition of} the surface sag ~\cite{Walker2009} and the normals and then {setting up} a procedure of ray-tracing at any point of this surface. Further, the diffraction efficiency can be computed by application of any numerical method. For a holographic freeform grating the recording and operation geometries vary significantly from one point to another, so the local gratings approach is used. 
Below we describe the approach to model the freeform surface in ray-tracing software, which was implemented in a custom dll-library for Zemax. Also we present the diffraction efficiency modelling procedure. It was realized by a combination of Zemax macros and Matlab-based numerical diffraction solvers.  

\subsection{Freeform description}
A freeform surface can be described by different sets of polynomials ~\cite{Ye2017}~. In the present case, one of the orthonormal polynomials sets -–- Zernike or Legendre polynomials are used for the surface description.

\begin{figure}
\begin{center}
\begin{tabular}{c}
\includegraphics[width=0.8\linewidth]{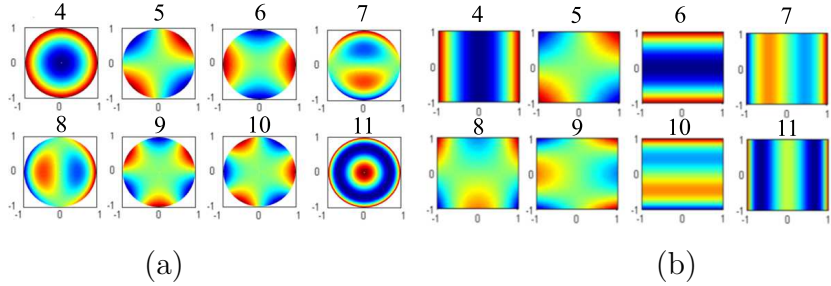}  
\\
(a) \hspace{6.0cm} (b)
\end{tabular}
\end{center}
\caption 
{ \label{fig:polynomials}
Modes of the orthonormal polynomials used for the freeform optical surfaces description: (a) standard Zernike sag polynomials and (b) Legendre polynomials.} 
\end{figure} 

The Zernike polynomials~\cite{Fleck2011} are widely used in optical engineering. They are orthonormal over the unity circle, which simplifies the freeforms design and optimization of them. Another polynomial set chosen for this task is the Legendre polynomials. They are orthonormal over the unity square.  Advantages of their use in some cases were demonstrated in a number of publications --- ref.~\cite{Ye2014,Ye2017}. To facilitate the further explanations we provide sag diagrams for the 4$^{th}$ to 11$^{th}$  modes of each set on Figure~\ref{fig:polynomials} (the piston, tip and tilt are the same, so they are excluded). It is necessary to mention that a freeform surface can be described by any of these polynomial sets, but due to some features of the numerical optimization process one way of the freeform description can be preferable~\cite{Muslimov2017}.

\subsection{Grating description}
For modelling of a holographic grating we used the general ray-tracing procedure~\cite{Welford75}. The vectorial equation used for computations is:

\begin{equation}
\label{eq:trace}
\vec{N}\times(\vec{r}_i-\vec{r}_d)=k \frac{\lambda}{\lambda_0}\vec{N}\times(\vec{r}_1-\vec{r}_2) \, ,
\end{equation}

\begin{figure}
\begin{center}
\begin{tabular}{c}
\includegraphics[height=5.5cm]{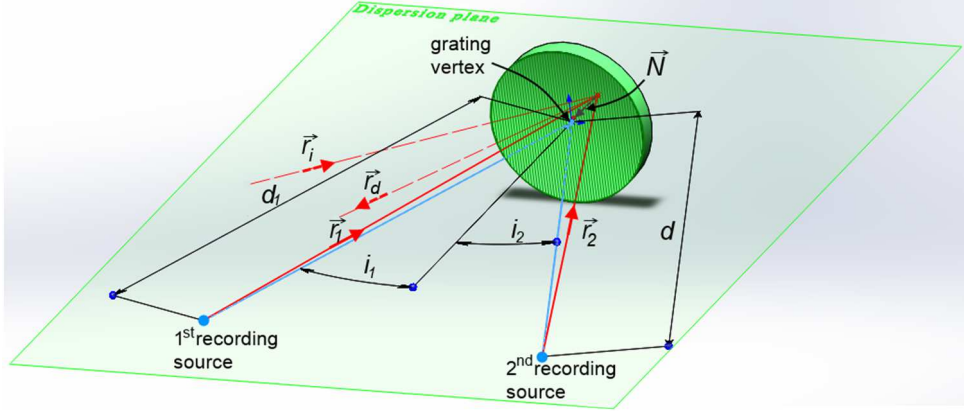}
\end{tabular}
\end{center}
\caption 
{ \label{fig:recording}
Definition of the holographic grating recording and operation geometry.} 
\end{figure}

Here  $\vec{N}$ is the normal vector, $\vec{r_i}$,$\vec{r_d}$  are vectors of the recording rays, $\vec{r_1}$,$\vec{r_2}$  are that of the incident and diffracted rays, $k$ is the order of diffraction, $\lambda_0$ and $\lambda$ are the recording and working wavelengths respectively.
Hereafter we assume that the grating is recorded by two point sources. The recording and operating geometry is presented on Figure~\ref{fig:recording}. {One should keep in mind that the recording sources coordinates are defined with respect to the grating vertex.}

In this case the grating works in reflection, but for a transmission grating the equation and all the definitions remain similar except for the sign before the diffracted ray vector and inclusion of the refraction index into equation~\ref{eq:trace}.
Combining the given descriptions of surface shape and grooves pattern, it is possible to create a user defined surface in an optical design software like Zemax (using customized .dll library) in order to use it with the standard ray-tracing and optimization tools.

\subsection{Diffraction efficiency modelling}

\begin{figure} [h]
\begin{center}
\begin{tabular}{c}
\includegraphics[width=0.75\linewidth]{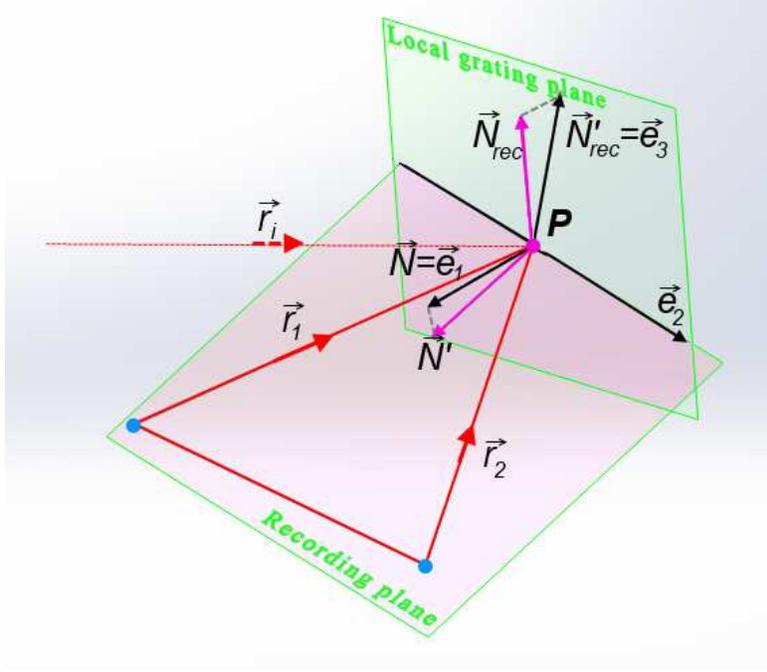}
\end{tabular}
\end{center}
\caption 
{ \label{fig:OE_local_grating}
The local grating approach: recording and operation geometry of the local plane classical grating.} 
\end{figure}

The diffraction efficiency depends strongly on the angle of incidence (AOI), groove orientation and groove profile. As it was mentioned, all of these parameters for a freeform holographic may vary significantly across the surface. So the diffraction efficiency is calculated for a number of local gratings. Each one of these gratings is supposed to be plane and have straight equidistant grooves. It is defined around each ray traced on the previous step by the local coordinate basis $(\vec{e_1},\vec{e_2},\vec{e_3})$, where $\vec{e_1}$ corresponds to the local surface normal and $\vec{e_3}$ gives the local grooves direction (see Figure~\ref{fig:OE_local_grating}).    

Then the modelling is performed for each of these local gratings separately using a precise numerical method - rigorous coupled wave analysis (RCWA). We consider two possible cases - a blazed surface-relief reflection grating and a volume-phase holographic (VPH) transmission grating. 

\begin{figure}[h]
\begin{center}
\begin{tabular}{c}
\includegraphics[width=0.75\linewidth]{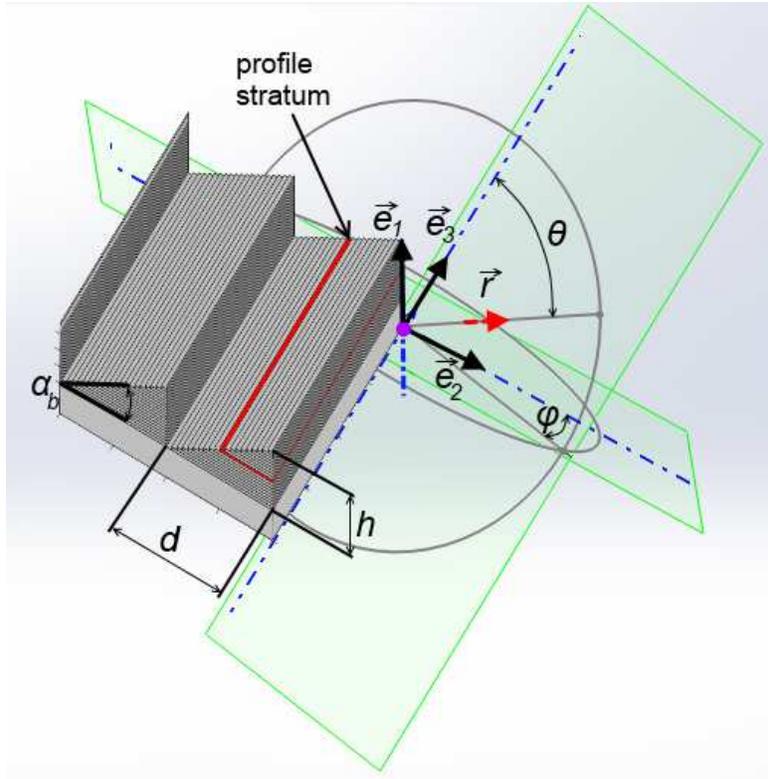}
\end{tabular}
\end{center}
\caption 
{ \label{fig:OE_local_grooves}
Modelling of the diffraction on blazed grooves of a local plane reflection grating.} 
\end{figure}

For the reflection grating case we used Matlab-based GD-calc software \cite{Johnson2006}. The grating profile there is represented by a number of layers or 'strata'. We assumed that the grooves are triangular (formed for instance by ion-beam etching). The grooves spacing $d$ is defined by the recording geometry:

\begin{equation}
\label{eq:freq}
N=\frac{1}{d}=\frac{sin(i'_1)-sin(i'_2)}{\lambda_0}\, ,
\end{equation}

where $i'_1$ and $i'_2$ are angles between the local normal $\vec{N}=\vec{e_1}$ and the projections of recording rays $\vec{r_1}$ and $\vec{r_2}$ to the { plane $\vec{e_1}\vec{e_2}$. They are different from those defined at the vertex, since the local recording plane does not coincide with the global tangential plane.}
The groove angle $\alpha_b$ corresponds to the blazing condition at the central wavelength of the working range $\lambda_c$ and the groove depth $h$ equals to $\frac{\lambda_c}{2}$. The incident ray vector is defined by two spherical angles as they are given in the GD-calc description. All the details are shown on Figure~\ref{fig:OE_local_grooves}.  

\begin{figure}[h]
\begin{center}
\begin{tabular}{c}
\includegraphics[width=0.75\linewidth]{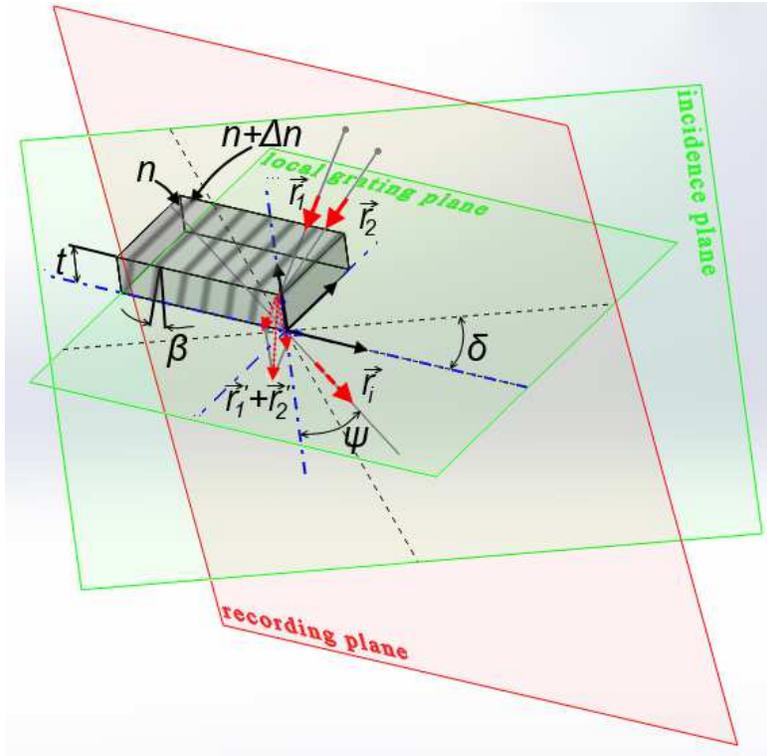}
\end{tabular}
\end{center}
\caption 
{ \label{fig:OE_local_VPH}
Modelling of the diffraction on fringes of a local plane transmission VPH grating.} 
\end{figure}

In the case of transmission VPH grating, the approach is similar. For the diffraction efficiency modelling another version of the RCWA method \cite{Chateau94} , which is implemented in \emph{reticolo} software, also operating in Matlab. Instead of grooves such a grating consists of fringes with modulated refraction index $n+\Delta n$. The main difference with the previous case is that the fringe inclination and therefore the diffraction efficiency is strictly dependant on the recording geometry. The inclination angle is defined as 

\begin{equation}
\label{eq:freq}
\beta=acos\frac{-\vec{N}\cdot(\vec{r'_1}+\vec{r'_2})}{|\vec{N}||(\vec{r'_1}+\vec{r'_2})|}\, ,
\end{equation}

where  $\vec{r'_1}$ and $\vec{r'_2}$ are the recording rays vectors refracted by the holographic media. 
In addition, the spherical angles $\psi$ and $\delta$ are re-defined according to the solver's requirements. 
The entire schematic is given in Figure~\ref{fig:OE_local_VPH}.

For both of the cases we developed sets of tools for the modelling. The surface shape and ray-tracing data was extracted from Zemax {using a macro}. Then it was readout by a Matlab code, which performs the efficiency computation for the entire surface and returns the average values as well as data on spatial distribution and polarization dependence.  

\section{Spectrograph with a reflection grating}
\label{sect:reflection} 
\subsection{Optical design}
\begin{figure}
\begin{center}
\begin{tabular}{c}
\includegraphics[width=0.8\linewidth]{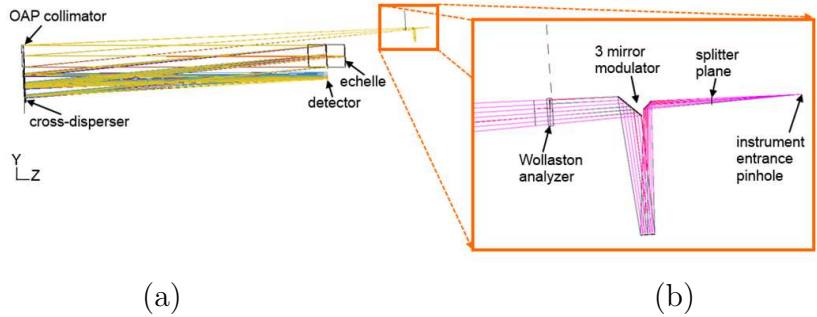}  
\\
(a) \hspace{6.0cm} (b)
\end{tabular}
\end{center}
\caption 
{ \label{fig:MUVdesign}
Optical design of the POLLUX instrument MUV channel:
 (a) general view, (b) zoom-in of the polarimetric unit.} 
\end{figure}

The primary application of the developed design tools is the optical design of POLLUX, a high resolution UV spectropolarimeter for the LUVOIR mission concept under study in the frame of the current NASA decadal survey. The target spectral resolving power of the instrument is 120 000 in the extended UV region 90-390 nm\cite{Muslimov2017}. Such a high resolution can be achieved in an instrument with 3 channels, when each of them represents an \textit{echelle} spectrograph. It was decided by the consortium to split the channels as follows: the far UV (FUV) channel operates from 90 to 124.5 nm and is fed by a flip mirror; the medium UV (MUV) channel works between 118.5 and 195 nm, while the near UV (NUV) channel works in the range 195-390 nm and they are separated by a dichroic splitter. It must be also mentioned, that the spectral length of a diffraction order should be 6 nm at least to avoid splitting of some broadened analytic lines.
The design solutions for the polarimetric units are different for each channel, whilst the spectral part is similar for all 3 channels. The MUV channel optics general view is shown as an example on Figure~\ref{fig:MUVdesign}. More details about the design can be found in \cite{Muslimov2018b} and \cite{Muslimov2018}.

\begin{figure}
\begin{center}
\begin{tabular}{c}
\includegraphics[width=0.7\linewidth]{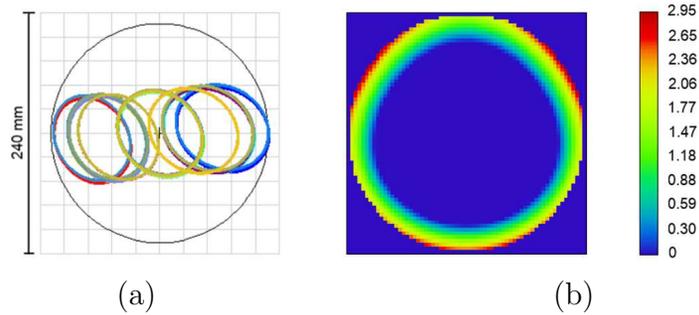}  
\\
(a) \hspace{5.0cm} (b)
\end{tabular}
\end{center}
\caption 
{ \label{fig:MUVsurface}
POLLUX MUV cross-disperser grating surface:
(a) footprint diagram and (b) map of the deviation from BFS in microns.
} 
\end{figure}

The entrance beam at F/20 passes through the pinhole and the polarimeter is collimated by an off-axis parabolic (OAP) mirror. Then it is dispersed by the \textit{echelle} grating. The cross-disperser represents a freeform holographic grating, so it simultaneously separates the echelle’s diffraction orders and focuses them on the detector. Use of such a complex element allows {correction of} the aberrations, achieves a high resolution and also decreases the total number of bounces and thus increases the throughput.  
One of the main features of this design is spatial separation of the spectral components on the cross-disperser’s surface (see Figure~\ref{fig:MUVsurface}, a). It means that the aberrations for this element considerably vary across the working range. So the grating properties and the surface shape should vary locally to compensate the aberrations. It also requires use of a freeform holographic grating. After the design and optimization, the grating frequency equals 212.3 $mm^{-1}$ and the recording sources rectangular coordinates are [100.194mm, 1913.946mm] and [-99.558mm, 1936.898mm] {if an  \emph{Ar} laser is used}. The focal length of the grating acting as a camera mirror is 1200 mm and its clear aperture is 215.4x98.3$mm^2$. The surface shape is described by the vertex sphere and six Zernike terms (primary coma, astigmatism and trefoil, $3^{rd}$ order spherical, astigmatism and quadrifoil). The deviation from the best-fit sphere (BFS) is shown on Figure~\ref{fig:MUVsurface}, b as a colormap. The root-mean square (RMS) deviation is 2.27 $\mu$m and the peak-to-valley (PTV) value is 3.36 $\mu$m.

\subsection{Image quality analysis}
\begin{figure}
\begin{center}
\begin{tabular}{c}
\includegraphics[width=0.7\linewidth]{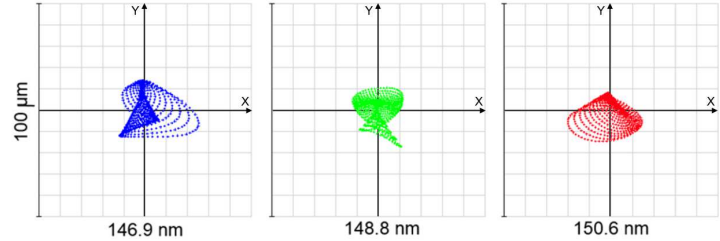}
\end{tabular}
\end{center}
\caption 
{ \label{fig:MUVspots}
Sample spot diagrams of the POLLUX MUV channel in the $40^{th}$ order.} 
\end{figure}

During the design and optimization process the image quality is assessed via the spot diagrams (see Figure~\ref{fig:MUVspots} for an example of diagrams for a single order). It is an easy way to {obtain} an image quality estimation. The designer should keep in mind that the minimized function is an RMS size of the spot and that the X (horizontal) direction corresponds to the \textit{echelle} dispersion direction, so the X size of the spot defines the spectral resolution.  Also it should be noted that the order spectral length is 3.7 nm without the part overlapping with the adjacent orders, though the full length on the detector corresponds to 6.1 nm. 

\begin{figure}
\begin{center}
\begin{tabular}{c}
\includegraphics[width=0.8\linewidth]{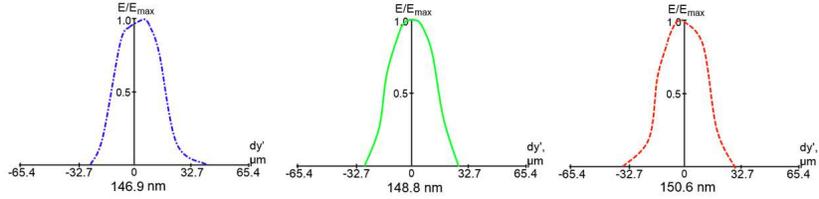}
\end{tabular}
\end{center}
\caption 
{ \label{fig:MUVif}
Sample instrument functions of the POLLUX MUV channel in the $40^{th}$ order (the entrance slit projection width is 31.2 $\mu$m).} 
\end{figure}

However, the dimensions of the spots do not correspond to the spectral resolving power directly. In order to compute the spectral resolving power the spectrograph’s instrument functions (IF) are calculated. The entrance slit width is 31.2 $\mu m$, which corresponds to $0.03''$ on sky. The results for the same control wavelengths in a single order are given in Figure~\ref{fig:MUVif}. {The slight asymmetry seen in the plots appears because of the residual aberrations, mainly the coma-type aberrations. However, they do not almost do not affect the spectral resolving power.}

The aberrations correction is good enough, as the FWHM (full width at half maximum) of IF equals to the slit width. The same data re-computed to spectral resolving power units is summarised in Table~\ref{tab:MUV}.   

\begin{table}[ht]
\caption{POLLUX MUV spectral resolving power summary.} 
\label{tab:MUV}
\begin{center}     
\begin{tabular}{|p{1.5cm}|p{1.5cm}|p{2cm}|p{2cm}|p{3cm}|} 
\hline
\rule[-1ex]{0pt}{3.5ex} Wavelength, nm &	Order &	Reciprocal linear dispersion, nm/mm	& Instrument function FWHM, $\mu$m &	Spectral resolving power \\
\hline\hline
\rule[-1ex]{0pt}{3.5ex}  188.9 &	31 &	0.047 &	31.2 &	128 909  \\
\hline
\rule[-1ex]{0pt}{3.5ex}  191.9 &       &		& 31.2   &	131 008   \\
\hline
\rule[-1ex]{0pt}{3.5ex}  195.0 &       &		& 31.95   &	133 106   \\
\hline
\rule[-1ex]{0pt}{3.5ex}  150.6 &    40   &	0.036 & 31.2   &	132 673   \\
\hline
\rule[-1ex]{0pt}{3.5ex}  148.8 &       &		& 31.2   &	131 035   \\
\hline
\rule[-1ex]{0pt}{3.5ex}  146.9 &       &		& 31.95   &	126 359   \\
\hline
\rule[-1ex]{0pt}{3.5ex}  120.2 &    50   & 0.029& 31.95   &	129 700   \\
\hline
\rule[-1ex]{0pt}{3.5ex}  119.0 &       &		& 31.2   &	131 502   \\
\hline
\rule[-1ex]{0pt}{3.5ex}  118.0 &       &		& 31.2   &	130 391   \\
\hline

\end{tabular}
\end{center}
\end{table}

The spectral resolving power is higher than the required value by ~10\%, that allows us to leave some margins for possible manufacturing errors and misalignments. 
In general, this modelling shows that with a freeform holographic grating it is possible to achieve the required spectrograph performance, while the grating recording scheme is feasible and the surface asphericity is small.  

\subsection{Diffraction efficiency analysis}
\begin{figure}[h]
\begin{center}
\begin{tabular}{c}
\includegraphics[width=0.75\linewidth]{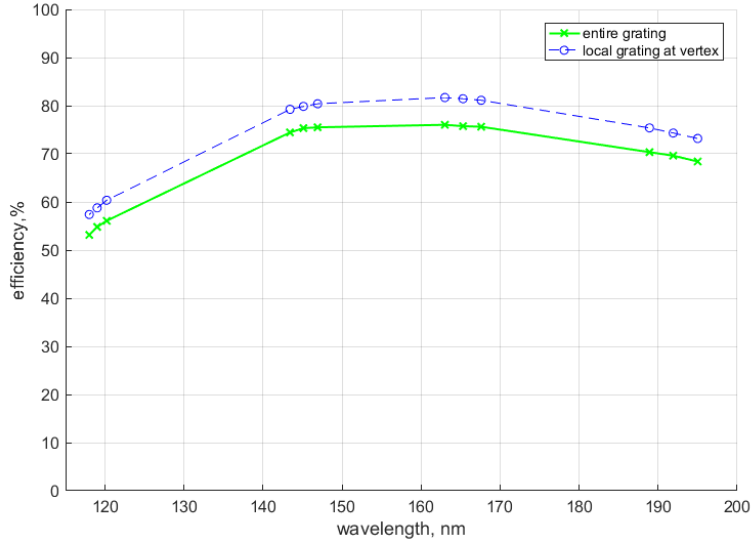}
\end{tabular}
\end{center}
\caption 
{ \label{fig:OE_DE}
Spectral dependence of the reflection freeform holographic grating diffraction efficiency.} 
\end{figure}

The diffraction efficiency was modelled with the technique described above. For simplicity we consider the pure spectral mode of POLLUX, i.e. exclude the influence of the polarimetric unit. Also it is supposed that the grating is made in Al with the permittivity of $(1.37+7.62i)^2$. For the instrument throughput budget, the obtained diffraction efficiency is multiplied by the coating reflectivity.  The main result of this modelling is the efficiency spectral dependence. The curve for unpolarized radiation is given in Figure~\ref{fig:OE_DE}. Also, for comparison the efficiency of a local grating around the surface vertex is presented. One can see that due to deviation from the optimal conditions across the surface, the overall efficiency is lower, while the curve shape is generally the same. {Comparison of the two curves indicates the effect of the local grating efficiencies variation consisting in a notable decrease of the overall efficiency. Thus , this variation should be always  taken into account when computing the efficiency of a holographic grating on a complex surface. In the future it may be also considered when optimizing the gratings parameters.} However, the {obtained} maximum value of 76.5\% with a decrease to 56.1\% at the edge is acceptable for the instrument.   

\begin{figure}[h]
\begin{center}
\begin{tabular}{c}
\includegraphics[width=0.75\linewidth]{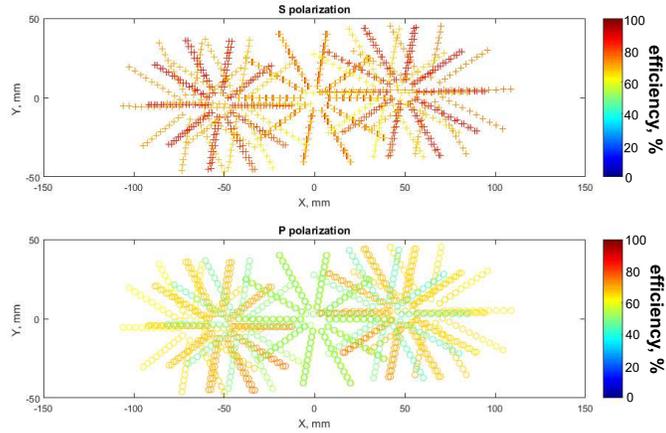}
\end{tabular}
\end{center}
\caption 
{ \label{fig:OE_pcolor}
Spatial distribution of the efficiency values over the reflection grating clear aperture for two polarization states.} 
\end{figure}

To estimate the efficiency change we investigated the values for local gratings at two polarization states. The result is shown in Figure~\ref{fig:OE_pcolor} as a colormap. It demonstrates that the grating introduces a notable polarization. In addition the efficiency values vary both spectrally and spatially, but since the grating is 'slow' {(i.e. it has a low surface steepness and operates with a high \textit{f}-number)} and the asphericity is small, the variation for a single wavelength is almost negligible. 

\begin{figure}[h]
\begin{center}
\begin{tabular}{c}
\includegraphics[width=0.75\linewidth]{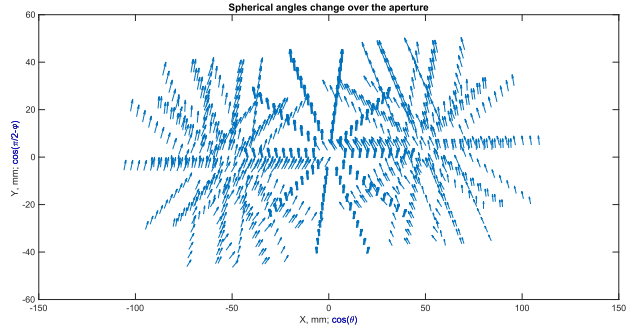}
\end{tabular}
\end{center}
\caption 
{ \label{fig:OE_vectors}
Variation of the incident ray angles over the reflection grating clear aperture.} 
\end{figure}

In order to illustrate the origin of the efficiency variation the spherical angles of the incident rays are plotted as a vector diagram in Figure~\ref{fig:OE_vectors}. The beginning point of each vector corresponds to the intersection point and the vector's coordinates correspond to the spherical angles (the trigonometric functions are used just for a better visualisation and have no special meaning). As the diagram shows, the incident ray vector coordinates change from one spectral component to another and slightly vary in each beam. {One must note that a similar variation occurs in the recording scheme as well. This explains the grating efficiency variation across the surface as shown in Figure~\ref{fig:OE_pcolor} qualitatively. But because of a large number of factors of different origins influencing the resultant DE, it becomes impossible to predict the efficiency variation theoretically. Instead the numerical modelling technique described above should be used.}

\section{Spectrograph with a transmission grating}
\label{sect:transmission} 
\subsection{Optical design}
\begin{figure}
\begin{center}
\begin{tabular}{c}
\includegraphics[height=8 cm]{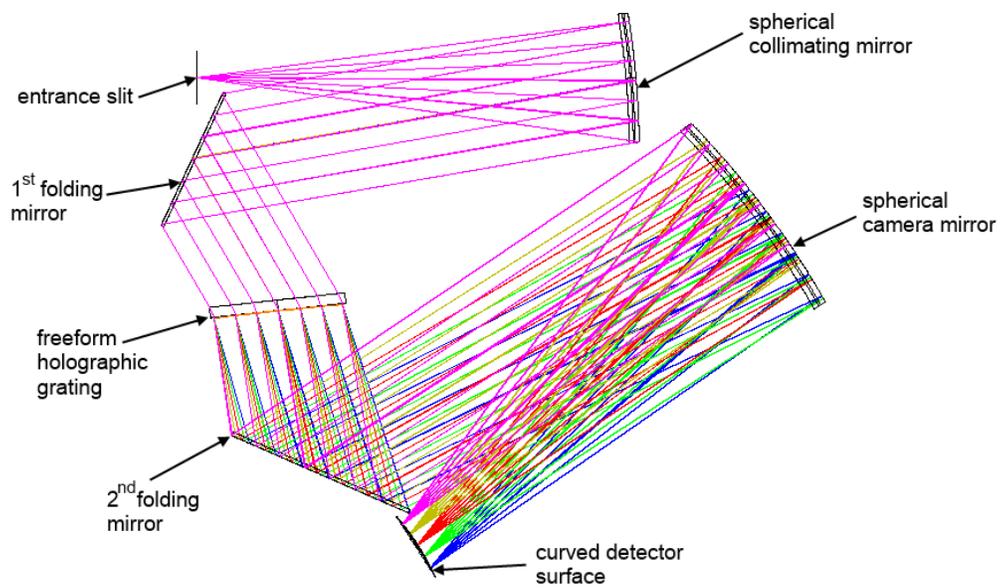}
\end{tabular}
\end{center}
\caption 
{ \label{fig:OE_virus_scheme}
Optical design of a double-Schmidt spectrograph with a transmission freeform holographic grating.} 
\end{figure}

The design and modelling tools developed for POLLUX can be generalized and applied to other optical systems. As an example we consider a spectrograph with a transmission freeform grating similar to the VIRUS instrument design\cite{Lee2010}. The considered design works in the visible domain 350-550 nm with F/3.1 aperture. Similarly to the prototype VIRUS instrument we start with a design using an inverted Schmidt telescope as the collimator and another Schmidt telescope as the camera. Focal distances of the both units are equal to 420 mm. In our case, the corrector plates for both of the Schmidt-type parts are united with the dispersive element thus forming the transmission freeform grating. Thanks to the Schmidt camera properties the spectrograph can operate with a long entrance slit up to 50 mm. Due to the use of an asymmetrical freeform surface it becomes possible to introduce tilts on the spherical mirrors and avoid any central obscuration in the optical scheme. The general view of the optical design is given in Figure \ref{fig:OE_virus_scheme}. Note that two flat folding mirrors were introduced to decrease the overall dimensions. We also should mention that in this case the focal plane is curved with a radius of 244.07 mm, though a similar result can be obtained with a flat image plane and a field flattener. 
An early version of this design was presented in \cite{Muslimov2018b} and we should emphasize that it was significantly changed. The main reason driving the modifications is the strong dependence of the  diffraction efficiency on the recording geometry, but we also improved the imaging performance and moderated the freeform deviation. 

\begin{figure}
\begin{center}
\begin{tabular}{c}
\includegraphics[height=5.5cm]{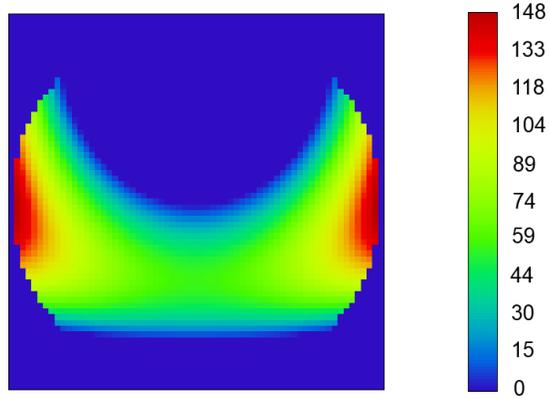}
\end{tabular}
\end{center}
\caption 
{ \label{fig:OE_virus_ff}
Deviation of the transmission freeform grating surface from the BFS in microns.} 
\end{figure}

The grating frequency at the vertex is 649 $mm^{-1}$ and the calculated recording sources coordinates for a Nd:YAG laser are (0.644 m, 9.268 m) and  (380.3 m, 870.0 m). Such sources practically correspond to two collimators with an {intentionally introduced defocus}. Since the aperture stop is located at the grating, its clear aperture is circular with a diameter of 124 mm. The optimization starts with a plane-parallel plate, but the vertex radius of curvature is set as a free parameter. Eleven Legendre modes, including all the modes up to the $4^{th}$ order symmetrical with respect to the YZ plane are used. The deviation from  the BFS of the surface is shown in Figure \ref{fig:OE_virus_ff}. The BFS radius is 2134.35 mm. The RMS deviation is 70.1$\mu$m and the PTV deviation is 177.4$\mu$m. These values are considerably larger than the ones obtained for previous design, however they are still manufacturable with the current level of technology. 

\subsection{Image quality analysis}
\begin{figure}
\begin{center}
\begin{tabular}{c}
\includegraphics[height=8cm]{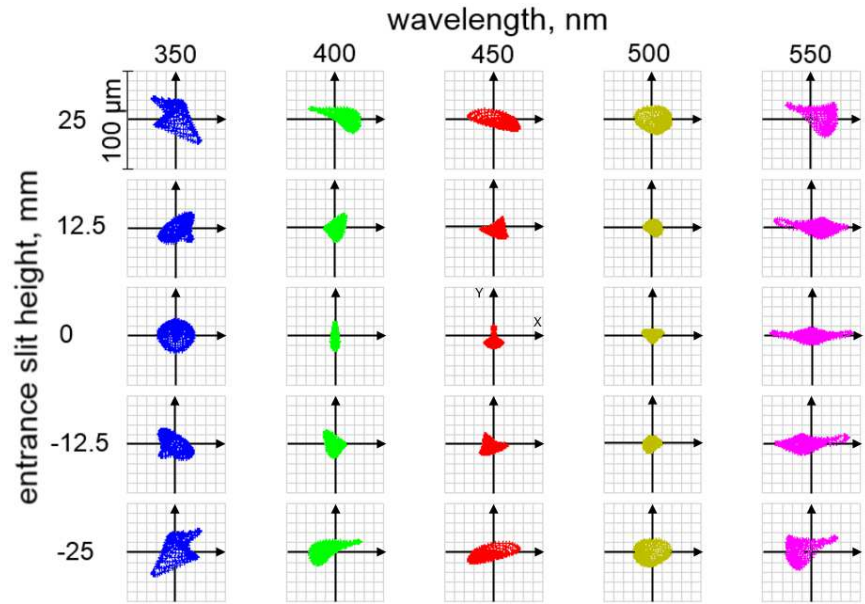}
\end{tabular}
\end{center}
\caption 
{ \label{fig:OE_virus_spots}
Spot diagrams of the double-Schmidt spectrograph for a long slit.} 
\end{figure}

Like for the reflective design, we consider the image quality estimations. The spot diagrams for long slit are presented on Figure~\ref{fig:OE_virus_spots}. {The horizontal \textit{X} axis corresponds to the slit height.} One can note that the aberrations are slowly growing towards the slit edges {due to the field aberrations, as it may be expected for any imaging system. But the main contribution to this spot blurring is made by astigmatism. It implies that the spot size in the dispersion direction \textit{Y} remains relatively stable, so the spectral resolving power does not change}.

\begin{figure}
\begin{center}
\begin{tabular}{c}
\includegraphics[width=0.8\linewidth]{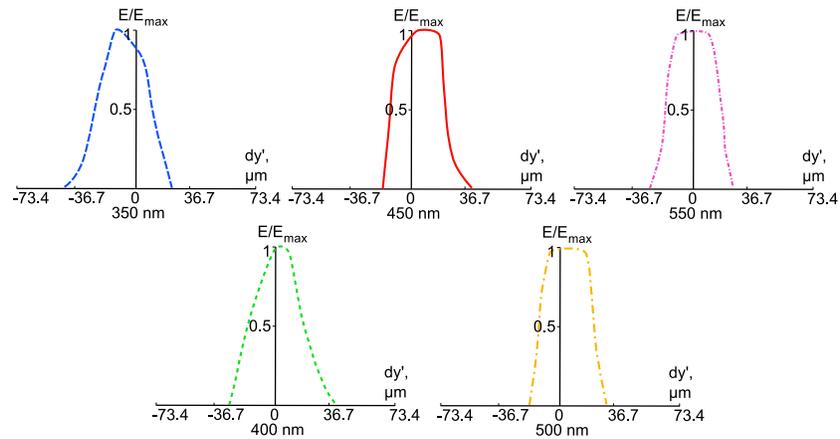}
\end{tabular}
\end{center}
\caption 
{ \label{fig:OE_virus_if}
Instrument functions of the double-Schmidt spectrograph (the entrance slit width is 35 $\mu$m).} 
\end{figure}

In order to prove and quantify the image quality we consider three spectrograph IFs. The IFs for the slit center are shown in Figure~\ref{fig:OE_virus_if}. The entrance slit width is equal to 35 $\mu$m in this case. The IF plots indicates the presence of a residual coma aberration, but the FWHM is very stable to the slit width across the entire working spectral range and for all the slit points.
 
Table~\ref{tab:VIR} summarises the spectral resolving power data. One can see a substantial gain in spectral resolution in comparison with the prototype design.

\begin{table}[ht]
\caption{The double-Schmidt spectrograph spectral resolving power summary.} 
\label{tab:VIR}
\begin{center}     
\begin{tabular}{|p{1.5cm}|p{1.5cm}|p{2cm}|p{2cm}|p{2cm}|} 
\hline
\rule[-1ex]{0pt}{3.5ex} Wavelength, nm & Slit height, mm &	Reciprocal linear dispersion, nm/mm	& Instrument function FWHM, $\mu$m &	Spectral resolving power \\
\hline\hline
\rule[-1ex]{0pt}{3.5ex}  350 &	   0 &	 4.0  &	35 &	2500  \\
\hline
\rule[-1ex]{0pt}{3.5ex}  400  &       &		& 35   &	2857   \\
\hline
\rule[-1ex]{0pt}{3.5ex}  450 &       &		& 35   &	3214  \\
\hline
\rule[-1ex]{0pt}{3.5ex}  500 &       &	   & 35   &	    3571  \\
\hline
\rule[-1ex]{0pt}{3.5ex}  550 &       &		& 35   &	3929   \\
\hline
\rule[-1ex]{0pt}{3.5ex}  350 &   25  &	 4.0& 37   &	2365   \\
\hline
\rule[-1ex]{0pt}{3.5ex}  400 &       &      & 35   &	2857   \\
\hline
\rule[-1ex]{0pt}{3.5ex}  450 &       &		& 35   &	3214   \\
\hline
\rule[-1ex]{0pt}{3.5ex}  500 &       &		& 35   &	3571   \\
\hline
\rule[-1ex]{0pt}{3.5ex}  550 &       &		& 35   &	3929   \\
\hline
\end{tabular}
\end{center}
\end{table}

\subsection{Diffraction efficiency analysis}

As it was mentioned before, in a case of VPH grating the recording geometry has an influence on both the imaging properties and diffraction efficiency. The initial design version\cite{Muslimov2018b} suffered from a low diffraction efficiency, since the recording source coordinates were optimized only accounting for the grating aberrations. It {compelled} us to re-optimize the entire design including deviation from the Bragg condition as a part of the merit function. 

\begin{equation}
\label{eq:freq}
cos(\beta-\theta)=\frac{N\lambda_c}{2n}\, ,
\end{equation}

where $\theta$ is the AOI and $n$ is the medium refraction index.
It is a relatively simple procedure, which can be done with standard software tools. 

After the design revision we computed the diffraction efficiency. We assumed that the holographic layer index is {equal to that} of the substrate, which is 1.46. For the modulation depth and layer thickness we used the optimization procedure described in \cite{Muslimov2013} considering only the local grating around  the vertex. The found values are $\delta n = 0.021$ and $t=10 \mu m$.  

\begin{figure}[h]
\begin{center}
\begin{tabular}{c}
\includegraphics[width=0.75\linewidth]{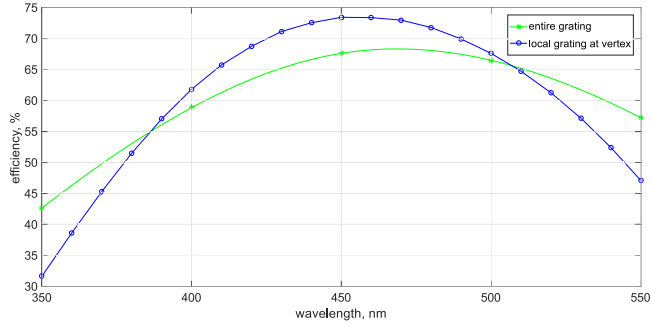}
\end{tabular}
\end{center}
\caption 
{ \label{fig:OE_virus_DE}
Spectral dependence of the transmission VPH freeform holographic grating diffraction efficiency.} 
\end{figure}

The modelling results are shown in Figure~\ref{fig:OE_virus_DE}. The efficiency reaches 68.3\% for the region center and decreases to 42.7\% at its edge. Again, for comparison, the vertex grating efficiency is plotted. The local grating is more effective at the central wavelength but also more selective. The curved flattening may be explained by the hologram reconstruction conditions variations across the surface.

\begin{figure}[h]
\begin{center}
\begin{tabular}{c}
\includegraphics[width=0.75\linewidth]{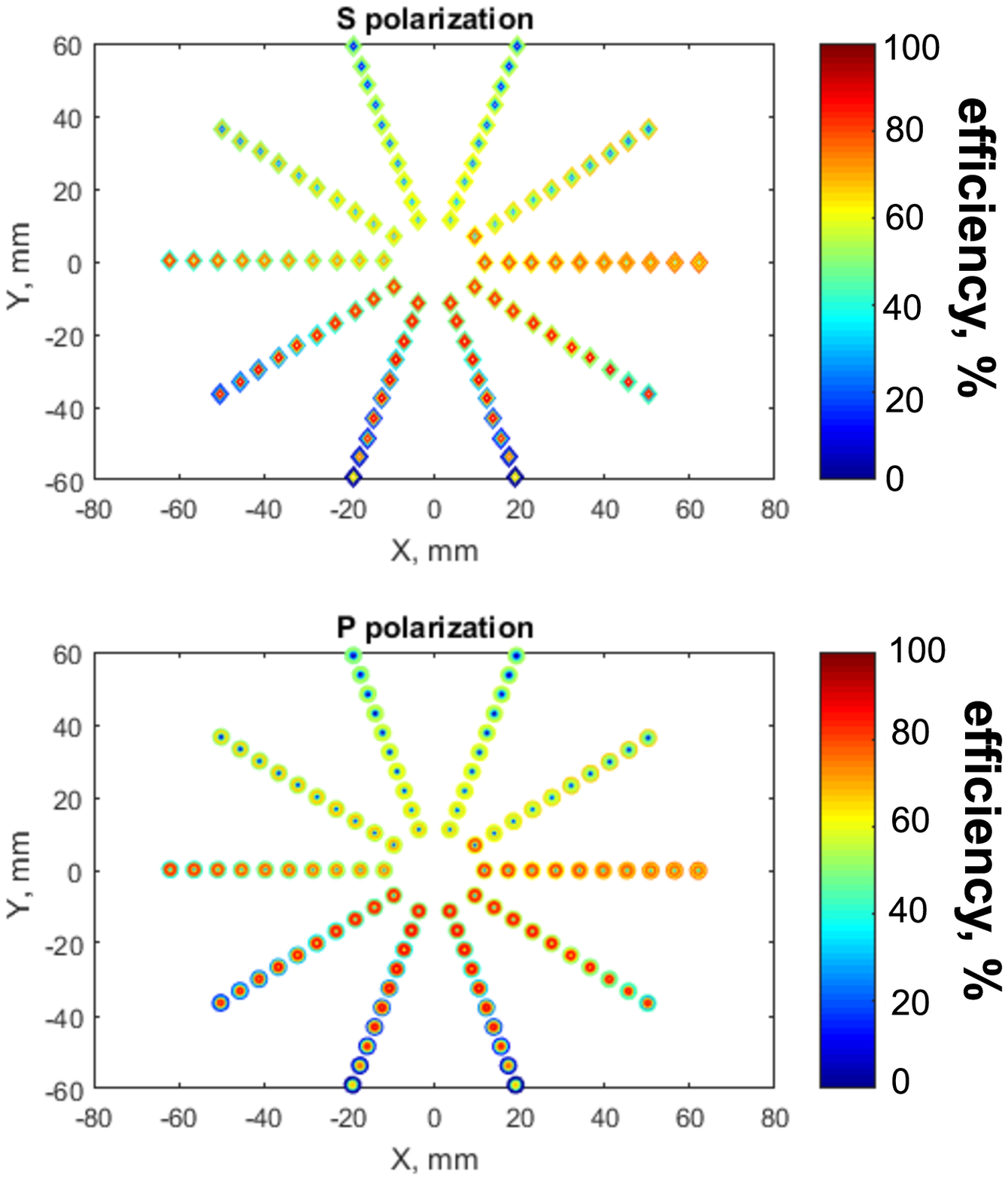}
\end{tabular}
\end{center}
\caption 
{ \label{fig:OE_virus_pcolor}
Spatial distribution of the efficiency values over the transmission VPH grating clear aperture for two polarization states.} 
\end{figure}

Similarly, we provide the efficiency variation map for two polarization states in Figure~\ref{fig:OE_virus_pcolor}. Because the spectral components are not separated at the grating surface, they are indicated with a different marker size. The diagram clearly shows the strong spectral and angular selectivity of the VPH grating. 

\begin{figure}[h]
\begin{center}
\begin{tabular}{c}
\includegraphics[width=0.75\linewidth]{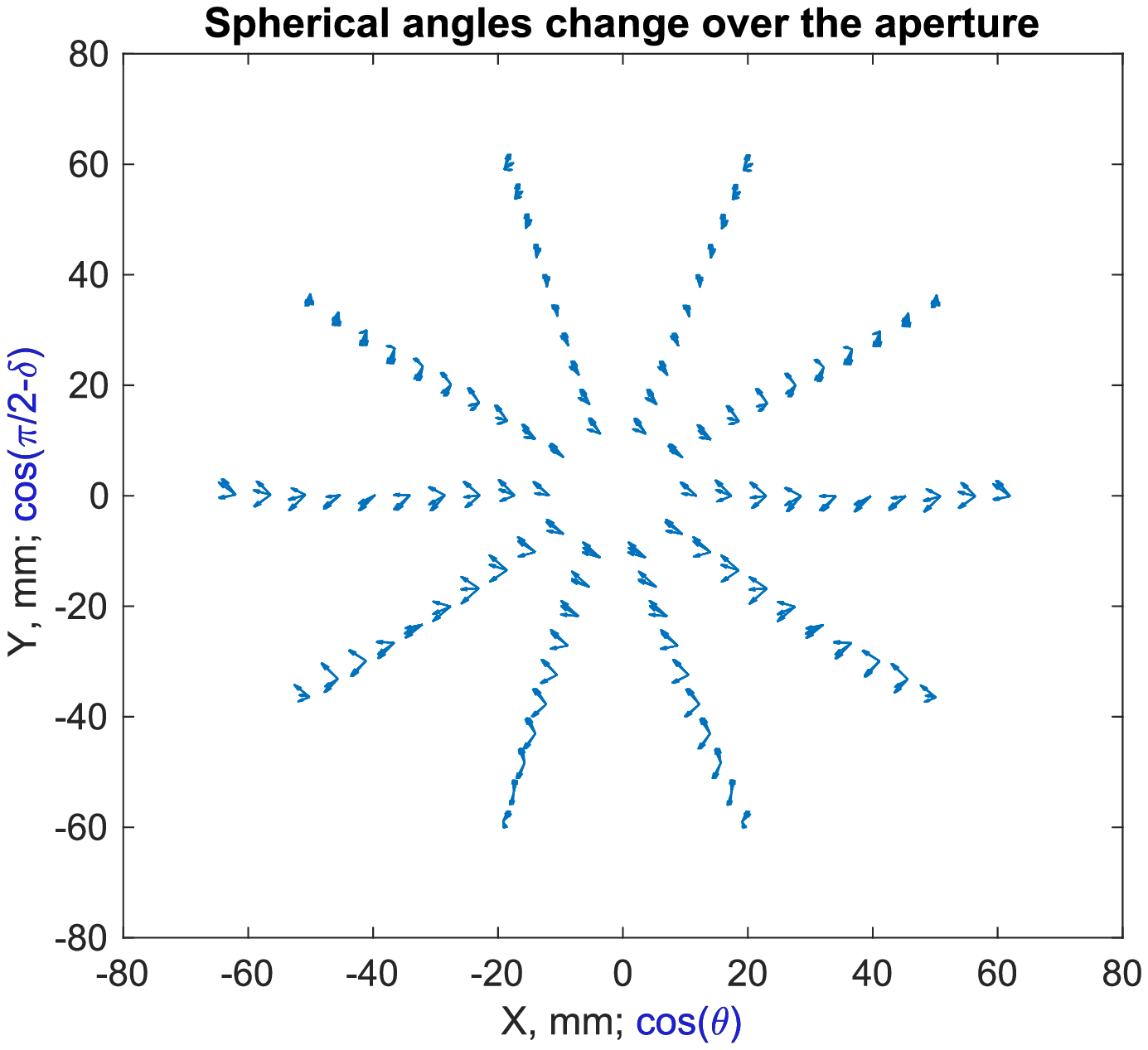}.
\end{tabular}
\end{center}
\caption 
{ \label{fig:OE_virus_vectors}
Variation of the incident ray angles over the transmission VPH grating clear aperture.} 
\end{figure}

The observed change of the efficiency can be explained by the AOI variation, which is illustrated by Figure~\ref{fig:OE_virus_vectors}. In comparison with the POLLUX design,  we have to include different field points (3 points in this case - the slit center and its edges). The figure shows that the hologram replay conditions {significantly change across the grating surface and the field of view. Moreover, their change pattern is non-symmetrical}.

\section{Conclusions}
\label{sect:conc}  
In the present paper we demonstrate the application of freeform holographic gratings for spectrographs design. This approach provides an outstanding possibility for aberration correction, which allows {improved performance and decreases} the number of optical elements. The first {example of an} application of this element is the optical scheme of the POLLUX spectropolarimeter for the LUVOIR mission. The freeform reflection grating works as the cross-disperser and camera simultaneously and allows the required spectral resolving power of $R>120 000$ {to be reached and minimizes} the number of bounces in the scheme. 
The second example is a demonstrative design of a double-Schmidt spectrograph with transmission grating. The design is notable for its absence of central obscuration and relatively high spectral resolving power reaching $R=3929$.
In both cases we demonstrated the influence of the recording and operation geometry on the grating efficiency. In the case of reflective blazed grating, the influence of grating shape and grooves pattern is limited by a moderate decrease of the grating efficiency. The efficiency curve remains relatively uniform and high. In the case of freeform VPH grating each of the local gratings has a strong angular and spectral selectivity. This effect is partially cancelled out by the hologram replay conditions variation. As a result, the efficiency spectral dependence becomes low and flat. Future research activities may be related to the optimization of the diffraction efficiency and the imaging performance simultaneously by means of combination of the ray-tracing and the local grating efficiency computation in a single software tool.         
In general, the design approach demonstrated here and the corresponding tools can be useful for development of future optical instruments with a special emphasis on advanced spectrographs for scientific research, e.g. for  astronomy.

\subsection{Disclosures}
The authors have nothing to disclose.

\subsection{Acknowledgments}

The authors acknowledge the support from the European Research council through the H2020 - ERC-STG-2015 - 678777 ICARUS program. 







\end{spacing}
\end{document}